\documentstyle[aps,prl,epsf,floats,preprint]{revtex}

\tightenlines
\begin{document}

\draft
\title{From localization to delocalization in the quantum Coulomb glass}
\author{Thomas Vojta$^*$, Frank Epperlein, Svetlana Kilina, and Michael Schreiber}
\address{Institut f\"{u}r Physik, Technische Universit\"{a}t,
D-09107 Chemnitz, F.\ R.\ Germany\\
{\tt $^*$Contact information: phone +49 371 531 3147, fax +49 371 531 3151,
   email vojta@physik.tu-chemnitz.de}}

\date{\today}
\maketitle

\begin{abstract}
We numerically investigate how electron-electron interactions influence the
transport properties of disordered electrons in two dimensions. Our study
is based on the quantum
Coulomb glass model appropriately generalized to include the spin
degrees of freedom. In order to obtain the low-energy properties of this model we
employ the Hartree-Fock based diagonalization, an efficient numerical method
similar to the configuration interaction approach in quantum chemistry.
We calculate the d.c. conductance by means of the Kubo-Greenwood formula and
pay particular attention to the spin degrees of freedom.
In agreement with earlier results we find that electron-electron interactions
can cause delocalization. For spinful electrons this delocalization is
significantly larger than for spinless electrons.

\end{abstract}

\section{Introduction}
\label{sec:I}
The discovery of a metal-insulator transition (MIT) in the two-dimensional
electron gas in Si-MOSFETs \cite{2dMIT} has induced renewed attention
to the transport properties of disordered electrons. This MIT is in conflict
with the theory of localization for {\em non-interacting} electrons which
predicts that all states are localized in 2D.
The electron density in the Si-MOSFETs
is very low which makes the electron-electron interaction particularly
important. Thus it is generally assumed that some type of interaction effect
is responsible for this MIT.
One of the most remarkable findings about the MIT in Si-MOSFETs is that an
in-plane magnetic field (which does not couple to the orbital motion of
the electrons) strongly suppresses the conducting phase \cite{magnet}.
This suggests that the spin degrees of freedom play an important role
for the transition.
A complete understanding has, however, not yet been obtained.
There have been attempts to explain the experiments based on the
perturbative renormalization group \cite{runaway}, non-perturbative effects
\cite{nonperturb}, or the transition being a superconductor-insulator
transition rather than a MIT \cite{SIT}.

In order to attack the problem of disordered interacting electrons
numerically we have developed \cite{Dointer,HFD}
an efficient method, the Hartree-Fock based diagonalization (HFD)
which is related to the quantum-chemical configuration interaction
approach. We have used this method to study the influence of
interactions on the conductance in one
\cite{Transportin}, two \cite{Dointer}, and three \cite{pils98} dimensions.
We found a delocalizing tendency of the interactions for strong disorder
but a localizing one for weak disorder. Similar results have been obtained
by means of the density-matrix renormalization group \cite{schmitt} in one
dimension and exact diagonalization in two dimensions \cite{benenti}.
Since in most of the numerical work in the literature spinless electrons
were considered, there are not many results about the importance of the spin degrees
of freedom.

In this work we address this question by generalizing the
HFD method to spinful electrons. We then use it study the
influence of the spin degrees of freedom on the Kubo-Greenwood
conductance.

\section{Model and method}
\label{sec:II}

The generic model for {\em spinless} interacting disordered electrons is the
quantum Coulomb glass \cite{qcg}. In this paper we use a straight-forward
generalization of the quantum Coulomb glass to spinful electrons.
It is defined on a regular
hypercubic lattice with $g=L^d$ ($d$ is the spatial dimensionality) sites
occupied by $N=N_\uparrow +N_\downarrow=2K g$ electrons ($0\!<\!K\!<\!1$).
To ensure charge
neutrality each lattice site carries a compensating positive charge of
$2Ke$. The Hamiltonian is given by
\begin{equation}
H =  -t  \sum_{\langle ij\rangle, \sigma} (c_{i\sigma}^\dagger c_{j\sigma}
      + h.c.) +
       \sum_{i,\sigma} \varphi_i  n_{i\sigma} + \frac{1}{2}\sum_{i\not=j,
       \sigma,\sigma'}(n_{i\sigma}-K)(n_{j\sigma'}-K)U_{ij}
       + U_H \sum_{i} n_{i\uparrow} n_{i\downarrow}
\label{eq:Hamiltonian}
\end{equation}
where $c_{i\sigma}^\dagger$ and $c_{i\sigma}$ are the creation and annihilation
operators at site $i$ and spin $\sigma$, and $\langle ij \rangle$ denotes all
pairs of nearest-neighbor sites. $t$ is the strength of the hopping term,
i.e., the kinetic energy, and $n_{i\sigma}$ is the occupation number of spin state
$\sigma$ at site $i$.
We parametrize the interaction $U_{ij} = e^2/r_{ij}$ by its value
$U$ between nearest-neighbor sites. The Coulomb repulsion between two electrons
at the same site is described by the Hubbard interaction $U_H$
The random potential values
$\varphi_i$ are chosen independently from a box distribution of width $2
W_0$ and zero mean. The boundary conditions are periodic and the Coulomb
interaction is treated in the minimum image convention
(which implies a cut-off at a distance of $L/2$).

A numerically exact solution of a quantum many-particle system requires the
diagonalization of a matrix whose dimension increases exponentially with
system size. This severely limits the possible sample sizes. In order to
overcome this problem we have developed the HFD method. The basic idea is
to work in a truncated Hilbert space consisting of the corresponding
Hartree-Fock (Slater) ground state and the low-lying excited Slater states.
For each disorder configuration three steps have to be performed: (i) find
the Hartree-Fock solution of the problem, (ii) determine the $B$ Slater
states with the lowest energies, and (iii) calculate and diagonalize the Hamilton matrix in the subspace spanned by these states. The number $B$ of
new basis states determines the quality of the approximation, reasonable
values have to be found empirically.

\section{Results}
\label{sec:III}

The conductance of a quantum many-particle system can be
obtained from linear-response theory. It is essentially determined by the
current-current correlation function of the ground state.
The real (dissipative) part of the conductance (in units of $e^2/h$)
is given by the Kubo-Greenwood formula \cite{K57},
\begin{equation}
 {\rm Re} ~ G^{xx}(\omega) = \frac {2 \pi^2}  {\omega} \sum_{\nu} |\langle 0 | j^x|\nu \rangle |^2
     \delta(\omega+E_0-E_{\nu})
\label{eq:kubo}
\end{equation}
where $\omega$ denotes the frequency.
$j^x$ is the $x$ component of the current operator and $\nu$ denotes the eigenstates
of the Hamiltonian. Eq.\ (\ref{eq:kubo}) describes an isolated system while
 in a real d.c.\ transport experiment
the sample is connected to contacts and leads. This results in a finite life time $\tau$
of the eigenstates leading to an inhomogeneous broadening $\gamma = \tau^{-1}$
of the $\delta$ functions in (\ref{eq:kubo}) \cite{datta}. To suppress
the discreteness of the spectrum of a finite system, $\gamma$ should be
at least of the order of the single-particle level spacing. For our systems this requires
a comparatively large $\gamma \ge 0.05$. We tested different $\gamma$
and found that the conductance {\em values} depend on $\gamma$ but the
qualitative results do not.

\begin{figure*}
  \epsfxsize=15.5cm
  \centerline{\epsffile{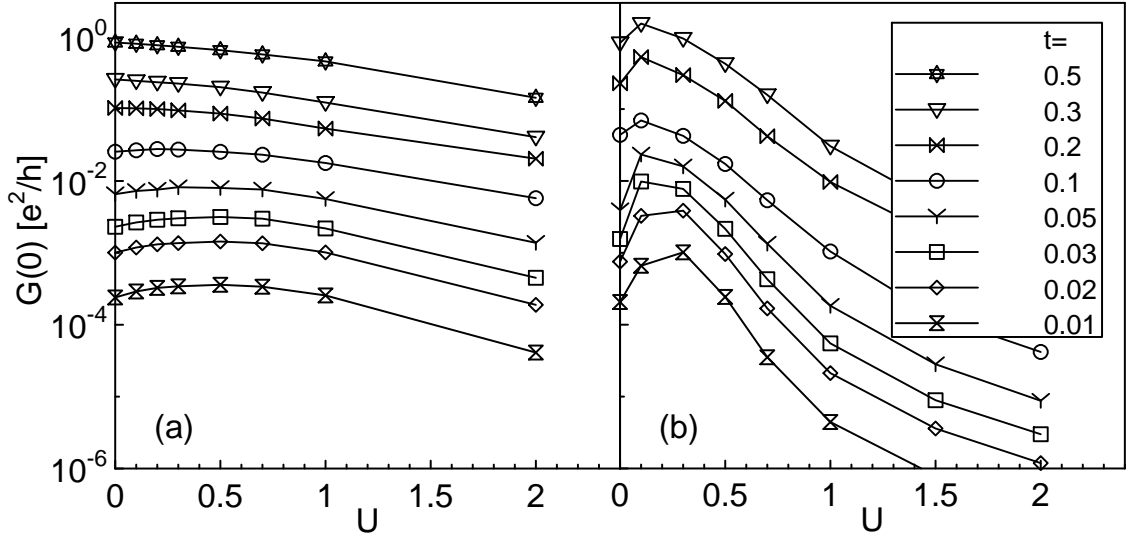}}
  \caption{d.c. conductance $G(0)$ for a system of
      $4 \times 4$ lattice sites occupied by (a)  8 spinless electrons
      or (b) 8 spin-up and 8 spin-down electrons for different
      $U$ and $t$. The disorder  strength is fixed to $W_0=1$, the Hubbard
      energy is $U_H=0.5$,
      the broadening is $\gamma=0.0625$, and the HFD basis size is $B=500$.
      The data points represent logarithmical averages over 400 samples.}
  \label{Fig:conductance}
\end{figure*}

The main results of this paper are summarized in Fig.\
\ref{Fig:conductance} which shows the disorder and interaction
dependence of the typical conductance for both spinless and spinful
electrons on a two-dimensional lattice of $4 \times 4$ sites.
The qualitative behavior in both cases is similar: In the strongly
localized regime (small $t$) a moderate interaction delocalizes the electrons
while a sufficiently strong interaction always strongly suppress the
conductance. This is the precursor of a Wigner crystal or Wigner glass.
With decreasing disorder (increasing $t$)
the interaction-induced enhancement of the
conductance also decreases and eventually vanishes.
The behavior of the conductance
can be attributed to the competition of two effects: First, the
interactions destroy
the phase of the electrons and thus the interference necessary for
localization. This is particularly effective if the localization length is
small to begin with. Second, the interactions introduce an additional source
of randomness which tends to increase the localization.

A comparison of the cases of spinless and spinful electrons  shows that
the interaction induced delocalization is significantly larger for spinful
electrons. Moreover, the enhancement seems to vanish at a larger kinetic
energy (which we did not reach in the simulations). A systematic  investigation
of the dependence of the conductance on $U$ and $U_H$ will be published
elsewhere.

In summary, we have studied the influence of electron-electron interactions
on Anderson localization for spinless and spinful electrons in two dimensions.
For strong disorder moderate interactions significantly enhance the transport.
This enhancement is much stronger for spinful than for spinless
electrons. Identifying a real phase transition and thus establishing a
connection between these findings and the experiments
on Si-MOSFETs requires a finite-size scaling analysis of the conductance.
This remains a task for the future.

\vspace*{0.25cm} {\small \noindent
This work was supported in part by the NSF under grant no.
DMR--98--70597 and by the DFG under grant no. SFB 393/C2.
T.V. thanks the Aspen Center for Physics and the University
of Oregon for hospitality during the completion of this paper.}

\end{document}